# Isospin-Based Supersymmetry in Neutron–Proton Pairing gap of fp-Shell Nuclei

Emami H.[1*], Sabri H.[2]

[1]Department of Physics and Energy Engineering, Amirkabir University of Technology, Tehran, Iran

[2]Faculty of Physics, University of Tabriz, Tabriz, Iran

**Abstract**

This study presents a systematic investigation of the effects of isospin, including both T=0 and T≠0 components, on nucleon pairing correlations in fp-shell nuclei. To this aim, the interacting boson model-4, which explicitly incorporates isospin and spin degrees of freedom, is employed to examine various mass formulas associated with neutron–proton correlations in atomic nuclei. Within this framework, neutron–proton pairing gaps are derived for the first time, providing a unified description of pairing correlations in even–even, odd–A, and odd–odd systems. Furthermore, to achieve a more realistic representation of local nuclear dynamics, effective asymmetric weighting coefficients are introduced into the conventional pairing-gap expressions, motivated by physical considerations related to shell structure, blocking effects, and isospin symmetry. These coefficients are determined through a fit to experimental binding-energy data and lead to a significant improvement in the agreement between theoretical predictions and observed systematics. Additionally, the results support the central assumption regarding the supersymmetry of the first fp-shell nuclei, highlighting the essential role of isospin-dependent effects in shaping neutron–proton pairing collectivity.

## 1. Introduction

The pairing in fp-shell nuclei is a dominant correlation that explains the even-odd binding energy differences, the ground-state spins, the low-lying collective excitations, and transitional behavior near shell closures (e.g., N=40). Nucleons (protons and neutrons) tend to form Cooper pairs in the nucleus due to the short-range attractive nuclear force. This pairing leads to the enhanced binding energy and therefore, pairing gap, a preference for even-even nuclei to have the $0^+$ state as the ground state, and also, in the odd-mass nuclei, often exhibit $1/2^+$ or $3/2^+$ ground states due to unpaired nucleons. The energy gaps, which are referred to as the difference in binding energies between even-even, odd-A, and odd-odd nuclei reflects pairing strength, the variation of the two-neutron separation energies ($S_{2n}$), which show even-odd staggering due to pairing, and also the low-lying $2^+$ states in even-even nuclei, which are collective vibrations influenced by pairing, are some of the nuclear phenomena that were described by using the effects of pairing gap in *fp* shell nuclei.

The investigation on the neutron-proton (n-p) pairing provides insights into the effects of isospin on the nuclear structures and is regarded as an important subject in recent years. Such studies make it possible to consider the connection of different isospin cases, e.g., isoscalar (T = 0) and isovector pairing (T = 1), among nucleons. There is no strong evidence for isoscalar pairing [1], even though the T=0 interaction is significantly stronger than the T=1 one [2]. To consider both types of pairing, including bosonic and boson-fermionic systems, such models are needed that are capable of accommodating either kind such as the supersymmetric model. Nuclear supersymmetry, conversely, is a theoretical framework that creates exact relationships between the spectroscopic characteristics of adjacent nuclei. The even-even and odd-odd nuclei function as composite bosonic systems, whereas odd-A nuclei are classified as fermionic. In this setting, nuclear supersymmetry offers a theoretical approach where bosonic and fermionic systems are considered part of the same supermultiplet [3]. The odd nucleon has isospin 1/2, which couples to the isospins of the s and d bosons in three ways. (No similar coupling choice exists for angular momentum, as the s-bosons' angular momentum is zero)[4]. The microscopic analysis of n-p pairing in nuclear matter at T=0, using bare nucleon-nucleon interactions, reveals a notable energy gap of 12 MeV [5]. Our objective is to analyze how the n-p pairing gap varies with the mass number A in this framework. A method to investigate the arrangement of atomic nuclei, including the influence of two-nucleon correlations, is by conducting a detailed examination of the mass surface of these nuclei. This entails looking at their general behavior and localized variations. This type of analysis is beneficial because experimental assessments of nuclear masses are extremely precise, and the quantity of isotopes with such data is steadily growing. [6,7]. The n-p interaction has been employed in conjunction with the pairing gap formulas [8,9].

The polarization effect [10] in the nuclear medium is known to either increase or reduce the T=1 neutron-neutron (nn) pairing gap, depending on the surrounding nuclear conditions [11]. The pairing gap is an energy gap of about 1 MeV, which exists between the ground state and nearly degenerate states with spin and parity ($J^{\pi}$) values of $0^{+}$and $2^{+}, 4^{+}, 6^{+}$,..[12].The pairing gap comes from the fact that it takes energy to break these pairs of nucleons. The interaction between nucleons leads to pairing gaps. Nuclei with a larger pairing gap are more stable and have longer lifetimes, while nuclei with smaller pairing gaps are more likely to undergo nuclear reactions.

Mass relations provide insight into the interaction strength between nucleons based on the mass number A and the likelihood of occupancy in the subshells close to the Fermi energy. The mass relationships for n-p pairing exhibit significant diversity [13]. This paper explores the concepts behind different mass relations in the supersymmetric model that pertain to n-p correlations within the first fp-shell nuclei. We are following Ref. [7] to the calculation of the n-p pairing gap by utilizing any mass relation in connection with the n-p pairing gap.

## 2. Theoretical framework

The pairing gap is an important concept in nuclear structure physics, particularly for nuclei with nucleons filling the fp shell (which includes the orbitals $1p_{3/2}$, $0f_{5/2}$,$1p_{1/2}$, and $0g_{9/2}$ for protons or neutrons). The pairing gap arises due to nucleon-nucleon correlations, leading to enhanced binding energy in even-even nuclei compared to odd-A or odd-odd nuclei. Also, the fp shell nuclei exhibit strong pairing correlations due to the high degeneracy of orbitals. This gap is influenced by monopole effects and quadrupole deformations, especially near shell closures, e.g., Z=28, 40 & 50 and N=28, 40&50.

To calculate the n-p pairing energy in an odd-odd nucleus that contains T=0 and T≠0 and has a n-p pair added to the double-closed nuclei, one must evaluate the difference between the n-p separation energy in the nucleus and the separation energies of a neutron and a proton [14]

$$\Delta_{np}^{o-o}(\mathrm{N,Z}) = S_{np}(\mathrm{N,Z}) - \left[S_n(\mathrm{N,Z}-1) + S_p(\mathrm{N}-1,\mathrm{Z})\right]$$
$$= \mathrm{B(N,Z)} + \mathrm{B(N-1,Z-1)} - \mathrm{B(N-1,Z)} - \mathrm{B(N,Z-1)}. \quad (1)$$

B(N, Z) denotes the binding energy, and we utilize the binding energies derived from the supersymmetric model in this formula, as well as in other similar mass formulas that will be described later, to calculate the pairing gap. This relationship, proposed in reference [15] for both even and odd values of N and Z, has been widely applied in various studies [16,17]. In odd–odd nuclei, there is additionally the residual interaction between the unpaired neutron and proton; thus, by adding isospin contributions such as T=0 and T≠0 in binding energy calculation, we can show this degree of freedom in IBM-4 can be related to the n-p pairing gap.

For an even-even nucleus that includes two pairs of neutrons and protons, the idea of n-p pairing can be explained as the difference in separation energy needed to extract all four nucleons from the core compared to the separation energies of the neutron and proton pairs in the nuclei (N, Z−2) and (N−2, Z) [18]:

$$\Delta_{np}^{e-e}(\mathrm{N,Z}) = \frac{1}{4}\,[\mathrm{B(N-2,Z-2)} + \mathrm{B(N,Z)} - \mathrm{B(N-2,Z)} - \mathrm{B(N,Z-2)}]. \quad (2)$$

The variation in binding energies among four even-even nuclei was suggested as a measure of the n-p interaction energy in reference [19].

Furthermore, a modified version of the generalized Eq. (1) for various types of nuclei was presented in Ref. [20]. The n-p pairing gap for odd-A nuclei is presented in the following:

$$\Delta_{np}^{e-o}(\mathrm{N,Z}) = \frac{1}{2}\,[\mathrm{B(N,Z)} - \mathrm{B(N,Z-1)} - \mathrm{B(N-2,Z)} + \mathrm{B(N-2,Z-1)}]. \quad (3)$$

$$\Delta_{np}^{o-e}(\mathrm{N,Z}) = \frac{1}{2}\,[\mathrm{B(N,Z)} - \mathrm{B(N,Z-2)} - \mathrm{B(N-1,Z)} + \mathrm{B(N-1,Z-2)}]. \quad (4)$$

Binding energy parametrization based on the supersymmetric model makes it possible to clarify the structure of the mass relations obtained and to reveal their relationship with the n-p interaction. Then we should explore the binding energies quantities in the supersymmetric model at first, to obtain the n-p pairing gap quantities for the first half of the fp-shell nuclei.

Nuclear supersymmetry is a phenomenon involving composite particles that connects the features of bosonic and fermionic systems. This concept is contextualized within the Interacting Boson Model (IBM) of nuclear structure [21]. Among various models of nuclear structure, the IBM and its extensions have been exceptionally successful in offering a cohesive framework for even-even [21] and odd-A nuclei [22]. Several extensions of the IBM-1, referred to as IBM-2, IBM-3, and IBM-4, have been developed by adding additional degrees of freedom to the bosons. We have chosen IBM-4 bosons that possess angular momenta of 0 and 2, in addition to various spin and isospin quantum numbers [23-27]. The most comprehensive IBM Hamiltonian for even–even nuclei is represented as follows:

$$H_B = \sum_{b,c} \alpha_{b,c} B_b^{+} B_c + \sum_{b,c,d,f} \beta_{b,c,d,f} B_b^{+} B_c^{+} B_d B_f \quad (5)$$

In this context, $B_b^+(B_b)$ represents the creation (annihilation) operator for bosons, while $\alpha_{b,c}$, and $\beta_{b,c,d,f}$ denote phenomenological parameters that have interpretations related to single-particle energy and two-body matrix elements. To account for odd and odd-odd nuclei, unpaired fermions can be included, resulting in the interacting boson-fermion model (IBFM) [22]. The Hamiltonian is then extended to the following form:

$$H_{BF} = H_B + \sum_{i,j} \gamma_{i,j} A_i^+ A_j + \sum_{i,j,k,m} \delta_{i,j,k,m} A_i^+ A_j^+ A_k A_m + \sum_{b,c,k,i} \rho_{b,c,k,i} B_b^+ A_k^+ B_c A_i \quad (6)$$

The Eq. (6) is infrequently utilized, primarily due to an excess of phenomenological parameters. Where $A_k^+(A_k)$ represents the operator for creating (or annihilating) fermions. The operators that keep the total count of particles (both bosons and fermions) constant, specifically, i.e., $B_b^+ B_c$, $A_k^+ A_i$, $A_k^+ B_b$and $B_b^+ A_k$, constitute a basis for the unitary superalgebra, which can be regarded as the largest (initial) algebra within the analyzed chain. Nucleons occupy the single-particle energy levels j = 1/2, 3/2, and 5/2, and the state vectors for single-particle fermions cover the 24-dimensional space where the most general unitary transformation group U(24) has been defined. The spin value σ and the isospin value τ correspond to those of two nucleons in LS-coupling, specifically, σ = 0 and τ = 1, (σ = 1 and τ = 0). Then, the isospin quantum number as degrees of freedom can affect the pairing of a nucleus and especially in this case, the pairing gap. Therefore, the one-boson space has a dimension of 36, and U(36) represents the unitary boson transformation group, so U(36/24) can be introduced as an initial supersymmetry group. We have selected the following chain of groups:

$$\begin{aligned} U(36/24) &\supset U^B(36) \times U^F(24) \\ &\supset {U_L}^B(6) \times {SU_{ST}}^B(6) \times {U_L}^F(6) \times {SU_{ST}}^F(4) \\ &\supset {U_L}^{BF}(6) \times {SU_{ST}}^{BF}(4) \\ &\supset {SO_L}^{BF}(6) \times {SU_S}^{BF}(2) \times {SU_T}^{BF}(2) \\ &\supset {SU_J}^{BF}(2) \times {SU_{ST}}^{BF}(2). \end{aligned} \quad (7)$$

And the dynamical Casimir Hamiltonian according to Eq. (7) :

$$\begin{aligned} H = H_0 &+ Q\hat{C}_2[{U_L}^{BF}(6)] + W\hat{C}_2[{SO_L}^{BF}(6)] + R\hat{C}_2[{SO_L}^{BF}(5)] + Y\hat{C}_2[{SO_L}^{BF}(3)] \\ &+ L\hat{C}_2[{SU_S}^{BF}(2)] + F\hat{C}_2[{SU_J}^{BF}(2)] + Z\hat{C}_2[{SU_{ST}}^{BF}(2)]. \end{aligned} \quad (8)$$

The operators $\hat{C}_2$are the second-order Casimir operators, and the parameters Q, W, R, Y, F, and Z are phenomenological parameters. A key aspect of supersymmetry is that a single irreducible representation (IR) [23,26,28-30] of the unitary super-algebra includes states that possess varying quantities of bosons $N_B$ and fermions $N_F$. Following Ref. [23,30], we use the standard notation for IRs and quantum numbers labelling for the examination of binding energies in the supersymmetric model. In the supersymmetric framework, the total binding energy of a nucleus consists of the binding energy from a double magic nucleus, the bosonic vacuum added to the ground-state energy derived from Eq. (5) or (6). Thus, through Eq. (7), $H_0$ is represented in the following manner.

$$\begin{aligned} H_0 = {} & e_1\hat{C}_1[U(36/24)] + e_2\hat{C}_2[U(36/24)] + e_3\hat{C}_1[U^B(36)] + e_4\hat{C}_2[U^B(36)] \\ &+ e_5\hat{C}_1[U^F(24)] + e_6\hat{C}_2[U^F(24)] + e_7\hat{C}_2[{U_L}^B(6)] + e_8\hat{C}_2[{U_L}^F(6)] \\ &+ e_9\hat{C}_2[{SU_{ST}}^B(4)] + e_{10}\hat{C}_2[{SU_{ST}}^{BF}(4)]. \end{aligned} \quad (9)$$

For nuclei of the first half of the fp-shell, which have been studied in the IBM-3 and IBM-4 [25,31-32], we treat the supersymmetric model in the particle picture, the double magic $^{56}_{28}Ni$, and we will investigate only the supersymmetric model in the boson vacuum state. For a particle picture, the mass number [23]: $A = 56 + 2N_B + N_F = 56 + 2N - N_F$. The total number of particles, N, is given by the sum of the number of bosons, $N_B$, and the number of fermions, $N_F$. After calculations and definition of the binding energies constants in the supersymmetric model, we will investigate the n-p pairing gap using these formulas that contain the isospin quantum number (T) for nuclei which are located in the first half of the fp-shell, which are presented in the following:

(1) Odd-odd nuclei, T = 0

$$\begin{aligned} E_{gs} &= N(e_1 + 11e_2 + e_3 + 35e_4 + 5e_7 + 5Q + 4W) + N^2(e_2 + e_4 + e_7 + Q + W) \\ &\quad +5e_9 + 5e_{10} + 2B + 2F \\ &= \alpha N + \beta N^2 + \varepsilon. \end{aligned} \tag{10}$$

(2) Odd-odd nuclei, T ≠ 0

$$\begin{aligned} E_{gs} &= N(e_1 + 11e_2 + e_3 + 35e_4 + 3e_7 + 3Q + 2W) + N^2(e_2 + e_4 + e_7 + Q + W) \\ &\quad +T(Z + 4e_9 + 4e_{10}) + T^2(Z + e_9 + e_{10}) + 3e_9 + 3e_{10} + 2B + 2F \\ &= \alpha'' N + \beta N^2 + \gamma T + \delta T^2 + \varepsilon'. \end{aligned} \tag{11}$$

(3) Even-even nuclei

$$\begin{aligned} E_{gs} &= N(e_1 + 11e_2 + e_3 + 35e_4 + 5e_7 + 5Q + 4W) + N^2(e_2 + e_4 + e_7 + Q + W) \\ &\quad +T(Z + 4e_9 + 4e_{10}) + T^2(Z + e_9 + e_{10}) \\ &= \alpha N + \beta N^2 + \gamma T + \delta T^2. \end{aligned} \tag{12}$$

(4) Odd nuclei

$$\begin{aligned} E_{gs} &= N(e_1 + 11e_2 + e_3 + 33e_4 + 3e_7 + 5Q + 4W) + N^2(e_2 + e_4 + e_7 + Q + W) \\ &\quad +T(Z + 3e_9 + 4e_{10}) + T^2(Z + e_9 + e_{10}) \\ &\quad -e_3 - 34e_4 + e_5 + 24e_6 - 4e_7 + 6e_8 - \frac{7e_9}{4} + \frac{3e_{10}}{2} + \frac{3F}{4} + \frac{3L}{4} \\ &= \alpha' N + \beta N^2 + \gamma' T + \delta T^2 + \mu. \end{aligned} \tag{13}$$

The amount of $E_{gs}$ influences only the binding energies, leaving excitation energies and other spectra unaffected.

In order to account for the asymmetric contributions of neighboring nuclei arising from shell structure, isospin symmetry, and local pairing correlations, the standard n-p pairing-gap expressions are generalized by introducing effective weighting coefficients. These coefficients are allowed to differ for even–even, odd–A, and odd–odd nuclei, reflecting the distinct blocking effects and pairing configurations characterizing each nuclear class. Within this framework, the n-p pairing gap is expressed as a weighted combination of binding energies of neighboring systems, formulated explicitly in terms of neutron number N, isospin quantum number T, and mass differences. The resulting set of modified pairing-gap relations provides a flexible and physically motivated description of n-p correlations, enabling a refined comparison with experimental systematics across the fp-shell region.

For even–even nuclei, the weighted n-p pairing gap is given by:

$$\begin{aligned} N_F = 0, \quad & \Delta_{np}^{\mathrm{e-e}}(\mathrm{N,T}) \\ &= A_1 \mathrm{B}(\mathrm{N}-2, T) + C_1 \mathrm{B}(\mathrm{N,T}) + D_1 \mathrm{B}(\mathrm{N}-1, |T \pm 1|) + E_1 \mathrm{B}(\mathrm{N}-1, |T \mp 1|) + F_1. \end{aligned} \tag{14}$$

For odd–A nuclei, the corresponding expression reads

$$N_F = 1, \quad \Delta_{np}^{\mathrm{odd-nuclei}}(\mathrm{N,T})$$

$$= A_2 \mathrm{B}\left(\mathrm{N}-2, \left|T \pm \frac{1}{2}\right|\right) + C_2 \mathrm{B}(\mathrm{N},\mathrm{T}) + D_2 \mathrm{B}\left(\mathrm{N}-1, \left|T \mp \frac{1}{2}\right|\right) + E_2 \mathrm{B}(\mathrm{N}-1, |T \pm 1|) + F_2. \quad (15)$$

Finally, for odd–odd systems, the pairing gap is obtained as:

$N_F = 2, \quad \Delta_{np}^{o-o}(\mathrm{N},\mathrm{T})$

$$= A_3 \mathrm{B}(\mathrm{N}-2, T) + C_3 \mathrm{B}(\mathrm{N},\mathrm{T}) + D_3 \mathrm{B}\left(\mathrm{N}-1, \left|T \pm \frac{1}{2}\right|\right) + E_3 \mathrm{B}\left(\mathrm{N}-1, \left|T \mp \frac{1}{2}\right|\right) + F_3. \quad (16)$$

The introduction of asymmetric weighting coefficients in these expressions reflects the fact that neighboring nuclei contribute unequally to the effective n-p pairing field. Such asymmetry naturally arises from shell structure, blocking effects in odd systems, and isospin-dependent interactions, which modify the coherence of pairing correlations. Consequently, the weighted formulas provide a more realistic representation of local nuclear dynamics than the conventional symmetric mass-difference relations

The mentioned calculations in the method section are done for each nucleus that is located in the first half of the fp-shell, including Cu, Zn, Ge, As isotopic chains, and the constant parameters in the binding energies relationship in Eqs. (10-16) for each nucleus are obtained with the available experimental data (taken from [33]) using the Gaussian elimination method in Python. The constant parameters are shown in the Table1 and Table2.

## 3. Result and discussion

The binding energy can be expressed as an analytical function based on the number of nucleons [34]. Accurate estimation of the shell correction energy [35] is essential for the precise determination of binding energy, level density, and other structural properties of nuclear systems [36]. Observables such as nuclear masses, binding energy, and pairing gap can be used to characterize a nucleus and obtain information about nuclear correlations [37,38]. In the present analysis, the binding energies of fp-shell nuclei are first calculated within the isospin-based formalism and subsequently employed in the modified n-p pairing-gap expressions containing effective weighting coefficients. These coefficients are treated as phenomenological parameters and are determined through a least-squares fit to the available experimental mass data. No normalization constraints are imposed, since the coefficients represent effective coupling strengths rather than probabilistic weights. This procedure allows the relative physical importance of different neighboring configurations to be extracted directly from experimental systematics (see Table2), providing deeper insight into the local structure dependence of n-p pairing correlations.

Table1. Values of parameters (in keV) of mass formulae (10-13) for isotopic chains that are located in the first fp-shell region (Cu, Zn, Ge, As). Experimental data taken from [33].

| | $\alpha$ | $\alpha'$ | $\alpha''$ | $\beta$ | $\gamma$ | $\gamma'$ | $\delta$ | $\mu$ | $\varepsilon$ | $\varepsilon'$ |
|---|---|---|---|---|---|---|---|---|---|---|
| $56 \le A \le 68$ | -39.18 | -10.82 | -35.6 | 2.34 | -27.49 | 1.67 | 10.23 | -165.30 | -10.32 | -67.7 |

Examining Eqs. (10-13) reveals that appropriate combinations of coefficients can be treated as new phenomenological parameters. Interestingly, their quantity is somewhat lower than that of the original coefficients.

Table2. Constant coefficients of the n-p pairing gap formulae (14-16) for isotopic chains that are located in the first fp-shell region (Cu, Zn,Ge, As). Experimental data taken from [33].

| $\Delta_{np}^{e-e}(\mathbf{N},\mathbf{T})$ | $A_1$ | $C_1$ | $D_1$ | $E_1$ | $F_1$ |
|---|---|---|---|---|---|
| | -46.1 | 0.87 | 20.82 | 26.83 | 20119 |
| $\Delta_{np}^{odd-nuclei}(\mathbf{N},\mathbf{T})$ | $A_2$ | $C_2$ | $D_2$ | $E_2$ | $F_2$ |
| | 10.39 | 3.49 | -6.27 | -8.13 | 5481 |
| $\Delta_{np}^{o-o}(\mathbf{N},\mathbf{T})$ | $A_3$ | $C_3$ | $D_3$ | $E_3$ | $F_3$ |
| | 83.27 | -65.69 | -8.93 | -15.4 | 66261 |

The odd–even staggering assessment also includes nuclei with an odd mass number, necessitating blocking one pair of conjugate states, effectively removing it from the pairing scheme. Eqs. (1-4) relies on the binding energy, and the odd-even oscillation in binding energy relative to neutron number is recognized as one of the most reliable indicators of pairing in nuclei [39].

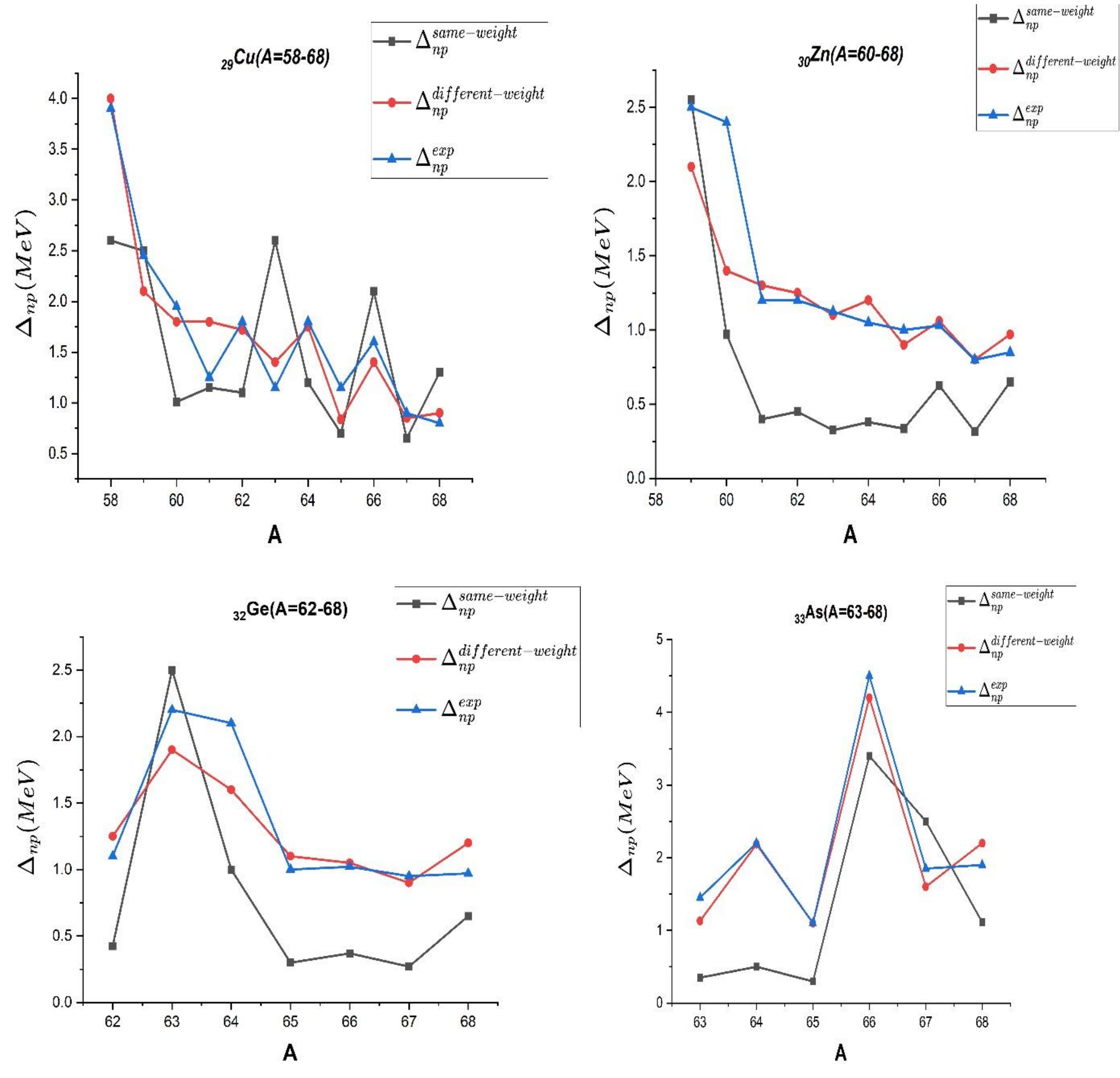

**Fig. 1**. n-p pairing gap ( $\Delta_{np}$) for the fp-shell nuclei as a function of the mass number (A), using supersymmetric model calculation from two perspectives of different and same weights and also experimental n-p pairing gap $\Delta_{np}^{exp}$ for similar isotopic chains that are located in the first fp-shell region (Cu, Zn, Ge, As). Experimental data are taken from Ref.[33].

The various n-p pairing gaps for each isotopic chain that is located in the first half of the fp-shell are shown in **Fig. 1**. As clearly seen in the comparison with experimental data, the introduction of asymmetric weighting coefficients leads to a substantial improvement in the reproduction of n-p pairing-gap systematics. This enhancement reflects the underlying physical reality that pairing correlations are strongly influenced by shell structure, proximity to closed shells, and isospin-dependent interactions. In particular, neighboring nuclei located across different shell regions or deformation regimes contribute unequally to the effective pairing field. The improved agreement obtained with the weighted formulas thus provides strong evidence that n-p pairing is governed by local structural properties rather than by simple geometric averaging, highlighting the essential role of isospin symmetry and shell effects in shaping pairing collectivity.

## 4. Conclusion

In this work, a comprehensive isospin-based analysis of n-p pairing correlations in fp-shell nuclei has been presented, aiming to clarify the interplay between shell structure, isospin symmetry, and local nuclear dynamics. By employing the IBM-4 framework, a unified description of pairing gaps across even–even, odd–A, and odd–odd systems has been achieved, enabling a consistent comparison with experimental mass systematics. A central outcome of this study is the demonstration that conventional mass-difference formulas, which implicitly assume symmetric contributions from neighboring nuclei, fail to capture essential structural asymmetries inherent in realistic nuclear systems. The introduction of effective, asymmetric weighting coefficients provides a physically motivated refinement, allowing the pairing gap to respond sensitively to variations in shell occupancy, blocking effects, and isospin-dependent interactions. The resulting improvement in the agreement with experimental data highlights the dominant role of local nuclear correlations in governing n-p pairing collectivity, beyond simplified geometric averaging schemes. The present findings underscore the importance of incorporating isospin degrees of freedom and symmetry considerations into phenomenological descriptions of nuclear pairing. They also lend further support to the applicability of supersymmetric concepts in the description of fp-shell nuclei, reinforcing the notion that collective and single-particle aspects of nuclear structure can be coherently unified within algebraic frameworks. As a natural extension of the current work, future studies will focus on exploring n-p pairing systematics in heavier mass regions, where the increasing role of deformation, continuum effects, and enhanced Coulomb interactions is expected to introduce new facets to pairing correlations. Such investigations will provide a broader perspective on the universality and limitations of isospin-based pairing descriptions across the nuclear chart.

**Acknowledgment**

This work is supported by the Research Council of the Amirkabir University of Technology and University of Tabriz.